\documentclass[conference]{IEEEtran}
\usepackage{cite}
\usepackage{amsmath,amssymb,amsfonts}
\usepackage{algorithmic}
\usepackage{graphicx}
\usepackage{textcomp}
\usepackage{xcolor}
\usepackage{tcolorbox}
\usepackage{booktabs} 
\usepackage{makecell, cellspace}
\usepackage{caption}
\usepackage{physics}
\usepackage{multirow}
\def\BibTeX{{\rm B\kern-.05em{\sc i\kern-.025em b}\kern-.08em
    T\kern-.1667em\lower.7ex\hbox{E}\kern-.125emX}}

\newcommand{\bt}[1]{\mbox{$\bf #1$}}
\def\l{\left(}
\def\r{\right)}
\newcommand{\img}{\bt x}
\newcommand{\cimg}{\hat{\bt x}}
\newcommand{\pixres}{\cimg(\theta) - \img}
\newcommand{\pixresi}{\cimg_i(\theta) - \img_i}
\newcommand{\jaco}{\bt J_{\sf f}}

\newcommand{\numpix}{n_p}
\newcommand{\numfeat}{n_{\sf f}}
\newcommand{\nummb}{n_b}

\DeclareMathOperator*{\argmin}{arg\,min}

\begin{document}
\title{Feature-Preserving Rate-Distortion Optimization in Image Coding for Machines\\
\thanks{Author email: samuelf9@usc.edu. SFM was also funded by the Fulbright Commission in Spain.}
}

\author{
\IEEEauthorblockN{Samuel Fernández-Menduiña, Eduardo Pavez, and Antonio Ortega}
\IEEEauthorblockA{\textit{Department of Electrical and Computer Engineering} \\
\textit{University of Southern California, Los Angeles, California, USA}\\}
}

\maketitle

\begin{abstract}
With the increasing number of images and videos consumed by computer vision algorithms, compression methods are evolving to consider both perceptual quality and performance in downstream tasks. Traditional codecs can tackle this problem by performing rate-distortion optimization (RDO) to minimize the distance at the output of a feature extractor. However, neural network non-linearities can make the rate-distortion landscape irregular, leading to reconstructions with poor visual quality even for high bit rates. Moreover, RDO decisions are made block-wise, while the feature extractor requires the whole image to exploit global information. In this paper, we address these limitations in three steps. First, we apply Taylor's expansion to the feature extractor, recasting the metric as an input-dependent squared error involving the Jacobian matrix of the neural network. Second, we make a localization assumption to compute the metric block-wise. Finally, we use randomized dimensionality reduction techniques to approximate the Jacobian. The resulting expression is monotonic with the rate and can be evaluated in the transform domain. Simulations with AVC show that our approach provides bit-rate savings while preserving accuracy in downstream tasks with less complexity than using the feature distance directly.
\end{abstract}

\section{Introduction}
Many images and videos are now primarily consumed by algorithms to extract semantic information. As a result, lossy compression methods are evolving to consider both human perception and computer vision performance \cite{zhang_call_2022, ascenso_jpeg_2023}, a framework often termed \emph{coding for machines} \cite{choi_scalable_2022}. While related ideas have been considered before \cite{ortega_compression_2000},  recent advances in solving computer vision problems with deep neural networks (DNN) \cite{lecun_deep_2015} have sparked renewed interest \cite{choi_deep_2018, choi_high_2018, le_learned_2021}. Approaches vary depending on the number of tasks and whether the encoder knows the task. For classification problems, where reconstructing the original content is unnecessary, algorithms based on the information bottleneck method \cite{tishby_information_2000} are sufficient. Similarly, if we consider a family of computer vision tasks, compressing the outputs of the first layers of a DNN \cite{choi_deep_2018}---which we will refer to as \emph{features}---and exploiting invariances \cite{dubois_lossy_2021, beferull_rotation_2003} yields substantial coding gains.

Instead, we focus on applications involving human supervision, which require the reconstruction of the image in addition to preserving performance on specific tasks, e.g., object detection and instance segmentation in video surveillance, traffic monitoring, or autonomous navigation \cite{jiang_adaptive_2023}. 
Both learned and traditional compression techniques can be used in this setting. Learned image compression (LIC) methods \cite{balle_end_2016} are popular because they can be trained with different distortion metrics \cite{ choi_scalable_2022}. However, these methods are complex \cite{ling_future_2022}, requiring millions of floating point
operations (FLOPs) per pixel \cite{guleryuz_sandwiched_2021}. 
Moreover, each encoder/decoder pair is optimized end-to-end for particular tasks \cite{le_image_2021, choi_scalable_2022} and may underperform on tasks outside its training scope.  

In contrast, traditional compression methods can adapt to different downstream tasks by parameter selection during encoding. In a coding for machines scenario,  
conventional distortion metrics, such as the sum of squared errors (SSE), must be complemented or replaced by task-specific losses. 
For example, the quantization parameter (QP)  can be tuned using importance maps derived from features \cite{choi_high_2018}. As another example, 
Fischer et al.~\cite{fischer_video_2020} propose a rate-distortion optimization (RDO) method to select QP and block partitioning, where the distortion metric is the distance between the 
outputs of a feature extractor obtained 
from the original image and a decoded image. 
As we argue in this work, minimizing the distance between features is particularly useful in \emph{transfer learning} scenarios---where the initial layers of a pre-trained DNN for a source task are used for different but related tasks---because the same encoder can be used for all the transferred applications.

\begin{figure}[t]
    \centering
    \includegraphics[width=\linewidth]{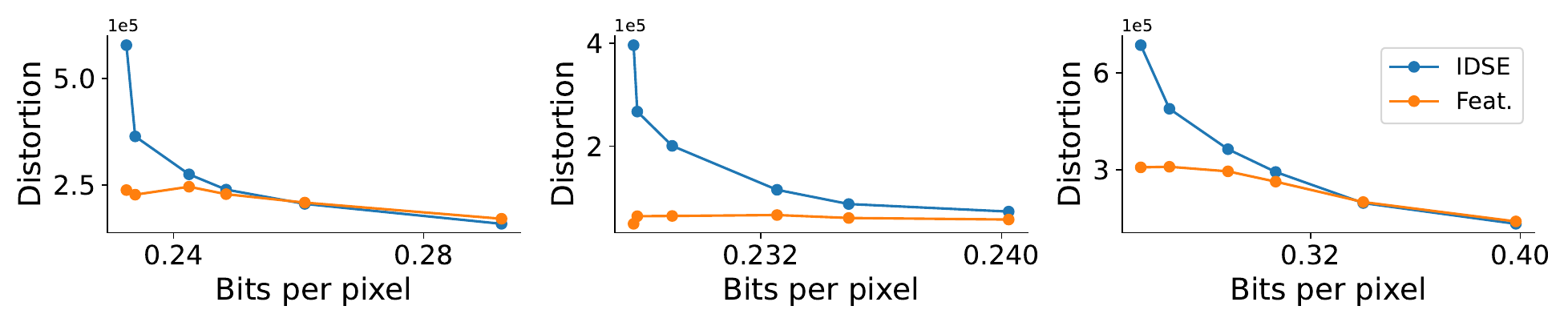}
    \caption{RD curves for the distance between features from VGG (Feat.) and the IDSE derived from the same feature extractor, for patches of size $128\times 128$ compressed using conventional AVC. The feature distance can be concave or non-monotonic with the bit rate. Only a small subset of operating points are reachable, compromising performance in terms of SSE. IDSE is quadratic, which leads to monotonic behavior.}
    \label{fig:examples_losses}
\end{figure}

Nonetheless, using the distance between features directly as a distortion metric in a conventional codec is problematic. In particular, neural network non-linearities often lead to concave or non-monotonic rate-distortion (RD) landscapes, so increasing rates may no longer reduce these distortions. 
Thus, the RD trade-off becomes harder to navigate. For instance, only a subset of low/high rate operating points may be reachable (cf.~Fig.~\ref{fig:examples_losses}), which may lead to reconstructions with a large SSE even for high rates, reducing visual quality. 
Moreover, while RDO decisions are made at the block level, the feature extractor requires the entire image to account for global context. Existing solutions \cite{fischer_video_2020} evaluate the feature extractor for each block, limiting access to global information. 
Furthermore, this approach may become computationally intensive \cite{gou_fast_2023} since, for each RDO candidate, it requires 1) a forward DNN pass, and 2) computing the distance in feature space, which often is higher-dimensional than the pixel space.

In this work, we propose a method to preserve important features for a set of tasks via RDO that overcomes these limitations. 
Our approach relies on three approximations. First, using Taylor's expansion, we approximate the loss by an \emph{input-dependent squared error} (IDSE)  involving the Jacobian matrix of the DNN with respect to the input image. Second, we localize the loss to evaluate it block-wise. Finally, since the dimensionality of the Jacobian increases the computational complexity, we estimate this matrix via metric-preserving dimensionality reduction  \cite{achlioptas_database_2003}. 
The resulting cost function can be evaluated in the transform domain using the transformed version of the Jacobian, and it can be combined with an SSE term so that RDO can address both visual quality and downstream tasks. 
Moreover, the loss is quadratic with the compression residual, making the RD curves monotonic.

\begin{figure}
\centering
\includegraphics[width=\linewidth]{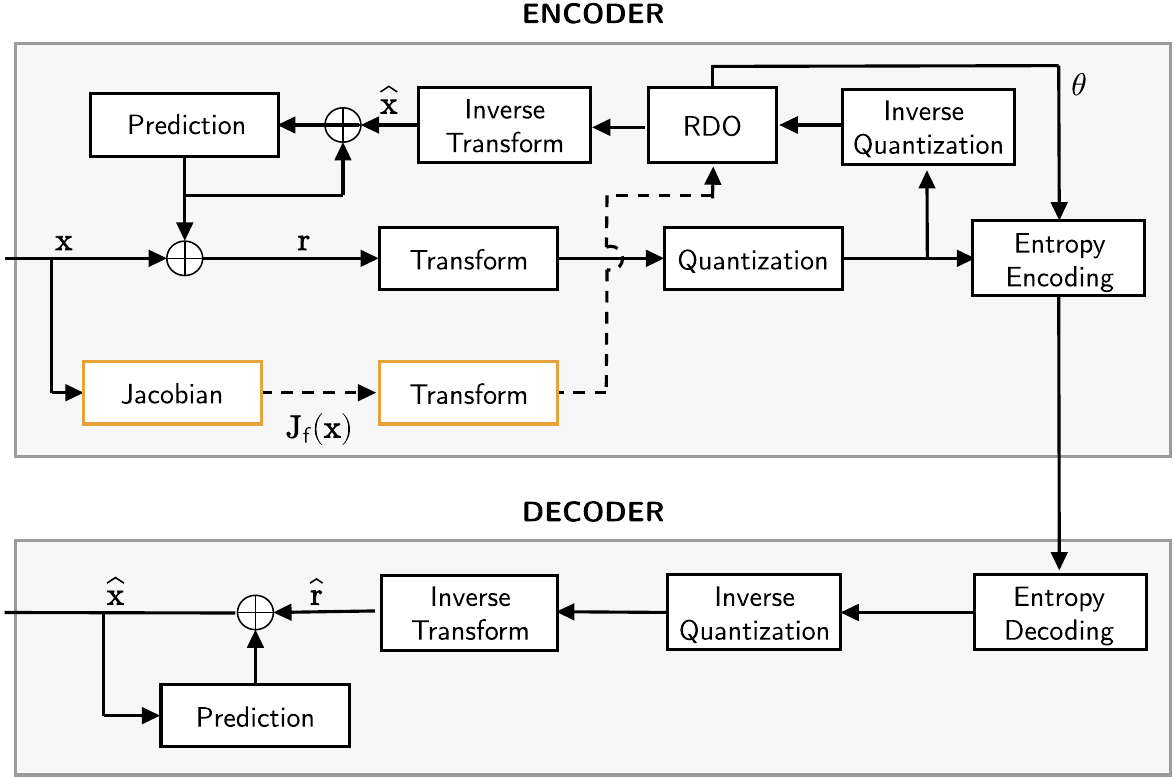}
\caption{Block diagram of the proposed codec, with the steps needed for feature-preserving RDO in yellow. Since we do not modify the decoder, it is compatible with standardized codecs.}
\label{fig:bd}
\end{figure}

Fig.~\ref{fig:bd} depicts the proposed codec, which is compatible with standardized decoders. The Jacobian is computed only once per image, regardless of the number of RDO candidates. While our setup is general and applicable to any RDO-based codec, we test it with AVC for selecting block partitioning. Even in this simple scenario,  IDSE-RDO provides more than 7\%  bit-rate savings in accuracy for 1) object detection/instance segmentation tasks in COCO 2017 dataset and 2) pedestrian detection/segmentation tasks in the PennFudan dataset \cite{wang_object_2007}.

\section{Preliminaries}
\textbf{Notation.} Uppercase bold letters, such as $\bt A$, denote
matrices. Lowercase bold letters, such as $\bt a$, denote vectors.
The $n$th entry of the vector $\bt a$ is $a_n$, and the $(i, j)$th entry of the matrix $\bt A$ is $A_{ij}$. Regular letters denote
scalar values.

\subsection{Rate-distortion optimization} Given an image $\bt x\in\mathbb{R}^{n_p}$ and its  $\nummb$ blocks, $\img_i$, $i = 1, \hdots, \nummb$, we aim to find parameters $\theta^\star$ from the set of possible operating points $\Theta$ satisfying \cite{everett_generalized_1963}:
\begin{equation}
    \theta^\star = \argmin_{\theta \in \Theta} \, d(\img, \cimg(\theta)) + \lambda \sum_{i = 1}^{n_b}\, r_i(\cimg_i(\theta)),
\end{equation}
where $d(\cdot, \cdot)$ is the distortion metric, $r_i(\cdot)$ is the rate for the $i$th coding unit, and $\lambda\geq 0$ is the Lagrange multiplier that controls the RD trade-off. We consider distortion metrics that are obtained as the sum of block-wise distortions,
\begin{equation}
\label{eq:local_gen}
d(\img_1, \hdots, \img_{\nummb}, \cimg_1(\theta), \hdots, \cimg_{\nummb}(\theta)) = \sum_{i = 1}^{\nummb} \, d_i(\img_i, \cimg_i(\theta)).
\end{equation}
This locality property, while true for SSE, does not hold for arbitrary metrics. Assuming that each coding unit can be optimized independently \cite{ortega_rate-distortion_1998}, we obtain
\begin{equation}
\label{eq:final_form}
  \theta_i^\star = \argmin_{\theta \in \Theta} \, d_i(\img_i,  \cimg_i(\theta)) + \lambda \, r_{i}(\cimg_i(\theta)), \quad i = 1, \hdots, \nummb,  
\end{equation}
where $\theta_i^\star$ are the optimal parameters for the $i$th block. This is the formulation of RDO we are concerned with in this work. A practical rule to control the RD trade-off \cite{wiegand_lagrange_2001} is 
\begin{equation}
\lambda = c \, 2^{(\mathrm{QP}-12) / 3},
\end{equation}
where $\mathrm{QP}$ is the quantization parameter, and $c$ varies with the type of frame and content \cite{ringis_disparity_2023}.

\begin{figure*}[t]
    \centering
    \includegraphics[width=\linewidth]{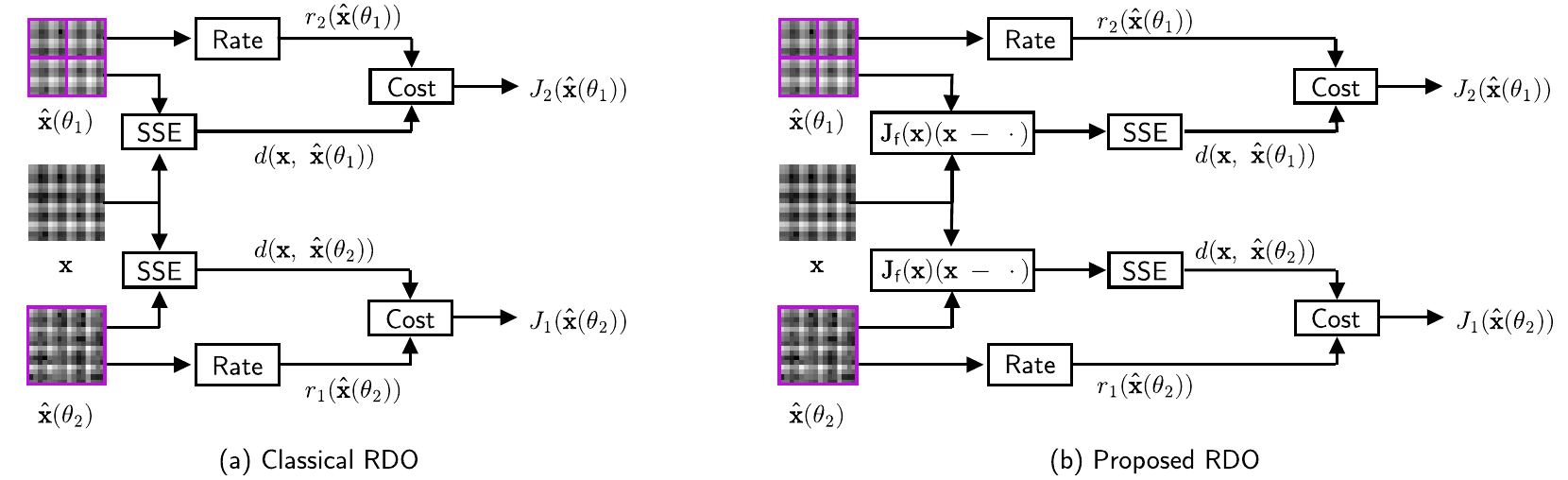}
    \caption{Process of comparing two RDO options, $\theta_1$ and $\theta_2$, using (a) classical SSE-RDO and (b) our proposed method. The main difference is the input-dependent squared error (IDSE) step using the Jacobian of the feature extractor, which encodes the information about the tasks of interest.}
    \label{fig:rdo_blocks}
\end{figure*}
\subsection{Feature extraction}
In this work, we focus on the output of a function $f(\cdot)$ comprising some of the layers of a DNN-based system, which we denote as the \emph{feature extractor}. We assume that the $f(\cdot)$ removes unnecessary information from the original image while preserving enough content to perform the task \cite{dubois_lossy_2021, beferull_rotation_2003}. By using a distortion metric based on the error introduced by compression on task-relevant features $\norm{f(\img) - f(\cimg)}_2^2$, we can maintain performance in computer vision problems. This setup is particularly relevant in transfer learning---where initial layers from a source task are used for a related application---because minimizing the distance at the output of a feature extractor preserves performance in all the transferred tasks. The next section explores how to preserve features via RDO.

\section{Feature-preserving RDO}
Given the feature extractor $f(\cdot)$, mapping images with $\numpix$ pixels to $\numfeat$-dimensional features, we aim to find:
\begin{equation}
    \theta^{\star} = \argmin_{\theta\in\Theta} \norm{f(\img) - f(\cimg(\theta))}_2^2 + \lambda \sum_{i = 1}^{\nummb}\, r_i(\cimg_i(\theta)).
\end{equation}
Note that this loss does not satisfy the locality property in Eq.~\eqref{eq:local_gen}. Existing methods \cite{fischer_video_2020} evaluate this task-dependent distortion by extracting features at the block level, which may become time-consuming if there are many RDO options to consider and hinders access to global information. In the following, we propose an alternative solution.

\subsection{Linearizing the feature extractor}
\label{sec:taylor}
We assume the feature extractor $f(\cdot)$ has second-order partial derivatives almost everywhere---a common assumption for analytical purposes \cite{jacot_neural_2018}. Let us define the Jacobian matrix of the network evaluated at the input image $\jaco(\img)\in\mathbb{R}^{\numfeat \times \numpix}$:
\begin{equation}
    J_{ij}(\img) = \pdv{f_i(\img)}{x_j}, \quad i= 1, \hdots, \numfeat, \ \ j = 1, \hdots, \numpix,
\end{equation}
that is, the derivative of the $i$th component of $f(\img)$ with respect to the $j$th component of $\img$. If we re-write $\cimg(\theta) = \img + (\cimg(\theta) - \img)$, we can apply Taylor's expansion to the feature extractor around $\img$:
\begin{equation}
    f(\cimg(\theta)) = f(\img) + \jaco(\img) (\pixres) + o(\norm{\pixres}_2^2),
\end{equation}
where $o(\norm{\pixres}_2^2)$ converges to zero at least as fast as $\norm{\pixres}_2^2$ when $\cimg(\theta) \to \img$. Under a high bit-rate assumption, compression errors are small, and we can keep only the first two terms:
\begin{equation}
\label{eq:jaco}
    \norm{f(\img) - f(\cimg(\theta))}_2^2 \approx  \norm{\jaco(\img)(\pixres)}_2^2.
\end{equation}
We refer to this loss as input-dependent squared error (IDSE). Therefore, the RDO problem can be written as
\begin{equation}
    \theta^{\star} = \argmin_{\theta\in\Theta} \,\norm{\jaco(\img)(\pixres)}_2^2 + \lambda \sum_{i = 1}^{\nummb}\, r_i(\cimg_i(\theta)).
\end{equation}
This optimization requires the whole image; thus, it is still unsuitable for making block-wise decisions. 
\subsection{Block-wise localization}
To facilitate the RDO process, we approximate $\jaco (\img)^\top \jaco(\img)$ by a block-diagonal matrix; intuitively, we assume the matrix is diagonally dominant, which mirrors local curvature approximations in the optimization literature \cite{schraudolph_fast_2002}. Then, if we denote the columns of the matrix $\bt J_{\sf f}(\img)$ corresponding to the pixels in the $i$th block by $\bt J^{(i)}_{\sf f}(\img)$, we obtain:
\begin{equation}
    \label{eq:localization}
    \norm{\jaco(\img)(\pixres)}_2^2 \approx \sum_{i = 1}^{\nummb} \,\norm{\jaco^{(i)}(\img)(\pixresi)}_2^2.
\end{equation}
Now, the RDO process can be split block-wise:
\begin{equation}
    \theta^{\star}_i = \argmin_{\theta\in\Theta} \, \norm{\jaco^{(i)}(\img)(\pixresi)}_2^2 + \lambda \, r_i(\cimg_i(\theta)),
\end{equation}
for $i = 1, \hdots, \nummb$, which has the form we stated in \eqref{eq:final_form}.
Fig.~\ref{fig:rdo_blocks} compares the proposed RDO method to SSE-RDO. 

\subsection{Jacobian approximation}
Computing the Jacobian is time-consuming: we need a backward pass for each entry in $f(\img)$. Also, IDSE requires the inner product of each row of the Jacobian with the error due to compression, $\pixresi$. 
We solve these problems by applying a metric-preserving linear transformation to $f(\img)$. Consider $h(f(\img)) = \bt Sf(\img)$, where $h(\img)$ is $\ell$-dimensional and $\ell \ll \numfeat$. Then, by the chain rule, 
\begin{equation}
    \bt J_{h\circ f}(\img) = \bt J_{h}(f(\img))\jaco(\img) = \bt S\jaco(\img).
\end{equation}
Thus, approximating the Jacobian reduces to $\ell$ backward passes and approximating IDSE to $\ell$ inner products. To preserve the metric, we rely on the Johnson–Lindenstrauss lemma \cite{achlioptas_database_2003}: given a set $X$ of $n_r+1$ points, with $n_r$ the number of RDO candidates, there exists a linear transformation $\bt S\in\mathbb{R}^{\ell\times \numfeat}$ such that, for all $\bt z, \bt y \in X$,
\begin{equation}
    (1-\epsilon)\norm{\bt z - \bt y}_2^2 \leq \norm{\bt S(\bt z - \bt y)}_2^2 \leq (1+\epsilon)\norm{\bt z - \bt y}_2^2,
\end{equation}
if $\ell > 8 \log(n_r)/\epsilon^2$. Different choices of $\bt S$, such as random Gaussian and Rademacher matrices  \cite{achlioptas_database_2003}, are explored in Sec.~\ref{sec:choose_s}. Under the localization assumption in Eq.~\eqref{eq:localization}, 
\begin{equation}
    \theta_i^\star = \argmin_{\theta\in\Theta} \,\norm{\bt S \jaco^{(i)}(\img)\l \pixresi\r}_2^2 + \lambda \, r_i(\cimg_i(\theta)),
\end{equation}
for $i = 1, \hdots, \nummb$. We denote this process as RDO-IDSE.

\subsection{Transform domain evaluation}
The IDSE can be computed in the transform domain. Given an orthogonal transform matrix $\bt D$, with $\bt y_i = \bt D^\top \bt x_i$, and denoting the quantized coefficients as $\hat{\bt y}_i(\theta)$, we can write
\begin{equation}
\theta_i^\star = \argmin_{\theta\in\Theta} \, \norm{\bt B^{(i)}(\img) (\bt y_i - \hat{\bt y}_i (\theta))}_2^2 + \lambda \, r_i(\hat{\bt y}_i(\theta)),
\end{equation}
for $i = 1, \hdots, \nummb$, where $\bt B^{(i)}(\img) = \bt S\jaco^{(i)}(\img)\bt D$. To extend it to separable transforms, let the $j$th row of $\bt B^{(i)}(\img)$ be $\bt b_j^{(i)}(\img)$, for $j = 1, \hdots, \ell$. Assume $\bt D$ separates between a row and a column transform such that $\bt D = \bt D_r \otimes \bt D_c$. If $\bt r_j^{(i)}(\img)$ is the $j$th row of $\bt S\jaco(\img)$ and $\bt R_j^{(i)}(\img)$ is its matrix version,
\begin{equation}
    \bt b_j^{(i)}(\img) = \bt r_j^{(i)}(\img)\bt D =  \mathrm{vec}(\bt D_c^\top \bt R_j^{(i)}(\img) \bt D_r)^\top,
\end{equation}
for $j = 1, \hdots, \ell, i = 1, \hdots, \nummb$, where $\mathrm{vec}(\cdot)$ is the vectorization operator; that is, we apply a block-wise transform to the matrix version of every row of the reduced Jacobian. 

 \begin{figure*}[th]
     \centering
     \includegraphics[width=\linewidth]{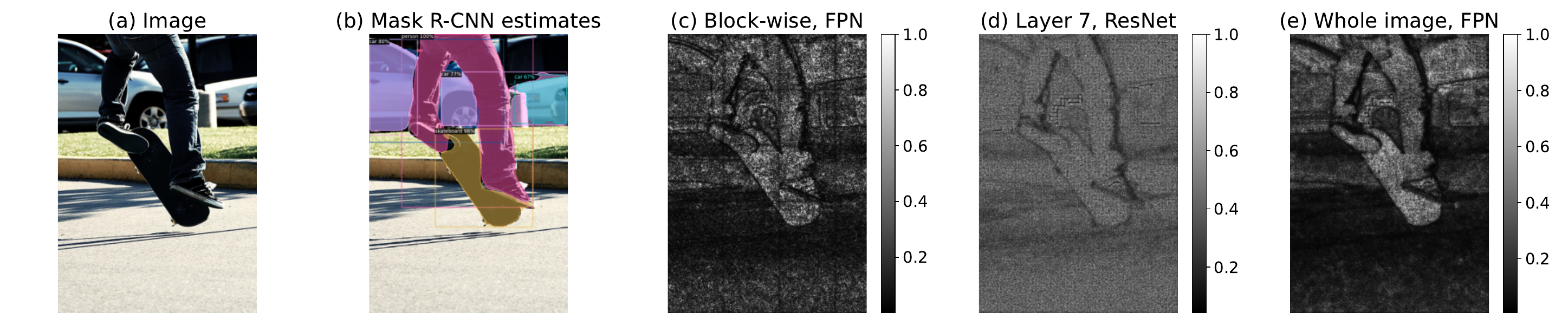}
     \caption{Image (a); Mask R-CNN estimates (b); and diagonal of $\bt J_{\sf f}(\img)^\top \bt J_{\sf f}(\img)$ (reshaped and scaled), obtained by (c) applying localization first and then expanding the metric block-wise, using blocks of size $128\times 128$ and an FPN; and expanding the metric directly with both (d) the first seven layers of the FPN's ResNet and (e) the whole FPN. Lighter regions receive more importance during RDO. Using the whole image and exploiting deep layers emphasizes relevant regions.}
     \label{fig:semantic}
 \end{figure*}
 
 \subsection{Combination with SSE}
To balance human and computer vision, the feature distance can be combined with SSE \cite{fischer_video_2020}. Our framework can also be combined with SSE, providing a pixel-level interpretation of the interaction between losses. In this section, we write $\cimg_i = \cimg_i(\theta)$. First, we expand the distortion term as
\begin{equation*}
    (\img_i - \cimg_i)^\top\jaco^{(i)}(\img)^\top\bt S^\top \bt S \jaco^{(i)}(\img)(\img_i - \cimg_i), \quad i = 1, \hdots, n_b.
\end{equation*}
Since SSE is the squared norm of the residuals,
\begin{equation}
\label{eq:balance}
(\img_i - \cimg_i)^\top\l \jaco^{(i)}(\img)^\top\bt S^\top \bt S\jaco^{(i)}(\img) + \tau \, \bt I\r (\img_i - \cimg_i),
\end{equation}
where $\tau\geq 0$ is the weight. The result is again an IDSE, but we apply Tikhonov regularization to the importance we give to each pixel. The larger $\tau$ is, the closer we are to SSE-RDO. If the matrix $\jaco(\img)^\top\bt S^\top \bt S\jaco(\img)$ were purely diagonal, $\tau$ would control the minimum SSE admissible for a given pixel. This regularization also ensures the weight matrix is full-rank. This is the loss we will consider in our experimental setup.

\subsection{Complexity}
\label{sec:complexity}
We provide the number of floating point operations (FLOPs) to evaluate the neural network; run-times with a real codec are given in Sec.~\ref{sec:choose_s}. We first compute the Jacobian, which requires a forward pass and $\ell$ backward passes---a backward pass having roughly twice the cost of a forward pass \cite{sepehri_hierarchical_2024}. To evaluate the network, we resize the images to the size used during training. An alternative to our method \cite{fischer_video_2020} is computing the feature distance block-wise, which requires evaluating the DNN and computing the distance of the features for each RDO candidate. In this case, no resizing is applied.

Assume the input has $h\times w$ pixels, and after resizing to compute the Jacobian, we get images of  $h'\times w'$ pixels; also, let $n_r$ be the number of RDO candidates. Let $C$ be the cost of the forward pass in terms of floating point operations per pixel (FLOPs/px). We use the same feature extractor for both approaches. Using the feature distance, we require $h\times w\times (n_r+1)\times C$ FLOPs to evaluate the cost throughout the image. We require  $h'\times w'\times (2\ell + 1)\times C$ FLOPs to sample the Jacobian. Assuming image sizes of $768\times 768$ pixels, resized images of size $224\times 224$, $n_r = 2$, and letting $\ell = 2$, our method reduces the number of FLOPs with respect to the approach that computes the feature distance by a factor of $7.06$.

\section{Empirical evaluation}
\label{sec:exper}
We consider object detection and instance segmentation using Mask R-CNN \cite{he_mask_2018}, with an FPN \cite{he_deep_2016} trained on COCO 2017 \cite{lin_microsoft_2014}. We focus on the COCO 2017 validation set and pedestrian detection/segmentation on the PennFudan dataset \cite{wang_object_2007} for feature transferability. We used an $8$-core CPU Intel Xeon-2640 and a GPU Nvidia Tesla P100 (16 GB VRAM).

\subsection{Semantic information}
Our method applies Taylor's expansion and then localizes the metric. 
An alternative is to localize the metric block-wise first---as suggested in \cite{fischer_video_2020}---and then apply Taylor's expansion as detailed in Sec.~\ref{sec:taylor}. However, evaluating the feature extractor block-wise hinders access to global semantic information. In Fig.~\ref{fig:semantic}, we show the diagonal of $\jaco(\img)^\top \jaco(\img)$ following both approaches and using the FPN as a feature extractor. We also argue that earlier layers might provide coarser information for the tasks of interest: we repeat the experiment above, evaluating the first seven layers of the ResNet 50 inside the FPN (Fig.~\ref{fig:semantic}--d). Obtaining the Jacobian in deeper layers with the whole image emphasizes the important regions for the tasks of interest.

\subsection{Compression experiments}
We use all-intra AVC baseline 4:2:0 \cite{sullivan_overview_2012}, but our method is compatible with any RDO-based codec. RDO chooses between $4\times 4$ and $16\times 16$ block partitions, evaluating the distortion on blocks of size $16\times 16$ pixels. We include an SSE term as in Eq.~\eqref{eq:balance} where $\tau$ equals the average Frobenius norm of $\jaco^{(i)}(\img)$. We adjust the Lagrange multiplier similarly to \cite{fischer_video_2020} but include the SSE in the normalization. We compress using $\mathrm{QP} \in    \lbrace 26, 28, 30, 32, 34, 36\rbrace$. We report the mean average precision (mAP@[0.5:0.05:0.95]) for each $\mathrm{QP}$.

\begin{figure}
    \centering
    \includegraphics[width=\linewidth]{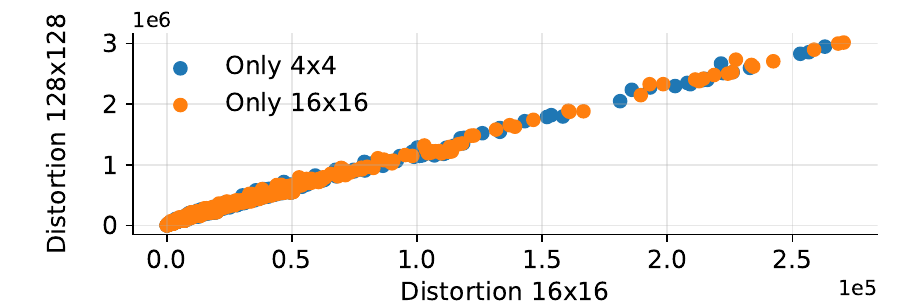}
	\caption{Feature distance with blocks of size  $128\times 128$ pixels (original metric \cite{fischer_video_2020}) and the corresponding $64$ sub-blocks of size $16\times 16$ (approximation). Images were compressed with AVC using only $4\times 4$ or $16\times 16$ modes to remove any mode decision effects. The Pearson correlation coefficient between metrics is $0.997$ for both mode decision setups.}    \label{fig:128vs16}
    \vspace{-1.5em}
\end{figure}

We also consider RDO with the distance between features (FD-RDO), which is inspired in \cite{fischer_video_2020}: we use the average of the Euclidean distance between features in the $5$th layer of VGG and the SSE. However, this setup was designed for block sizes of $128\times 128$ pixels while, due to codec and resolution constraints, we evaluate the distortion on blocks of size $16\times 16$. To assess this approximation, we evaluate the distance in feature space using blocks of $128\times 128$ (the original metric \cite{fischer_video_2020}) and the aggregate of the $64$ sub-blocks of $16\times 16$ (our approximation). We depict both quantities in Fig.~\ref{fig:128vs16}. Although the correlation is high,  the results we report may not represent the performance of the original method entirely. 

\subsubsection{COCO 2017 dataset}
\label{sec:choose_s}
We consider $200$ images from the validation set. For dimensionality reduction, we choose  $\bt S$ as (a) iid Rademacher, (b) iid Gaussian, and (c) DCT channel-wise, keeping the $16$ coefficients with the largest magnitude and reducing dimensionality using a Rademacher matrix. 

 We provide the average time to compute the Jacobian matrix in Table~\ref{tab:table_times}; the complexity scales proportionally to $\ell$ (cf. Sec.~\ref{sec:complexity}). We show the mAP-BD-rate saving \cite{bjontegaard_calculation_2001} with respect to SSE-RDO AVC (Table~\ref{tab:metrics_results}). Any setup performs better than SSE-RDO; in most cases, our method outperforms FD-RDO. For perceptual quality, we include Y-PSNR and Y-MS-SSIM \cite{wang_multiscale_2003}. IDSE-RDO gains slightly in MS-SSIM while FD-RDO does not; we conjecture that 1) feature distance has perceptual properties \cite{zhang_unreasonable_2018}, but 2) RD instabilities in FD-RDO lead to bad operating points for MS-SSIM, which IDSE-RDO avoids due to monotonicity (cf.~Fig.~\ref{fig:examples_losses}). We depict the RD curve for Rademacher sampling in Fig.~\ref{fig:coco_res} (a--b). To encode the luma channel, our approach with $\ell = 8$ and Rademacher sampling is $1.07$ times slower than AVC on average, while FD-RDO increases by a factor of $1.71$ with respect to AVC. 
 
\subsubsection{Pedestrian dataset}
\begin{figure*}
     \centering
     \includegraphics[width=0.49\linewidth]{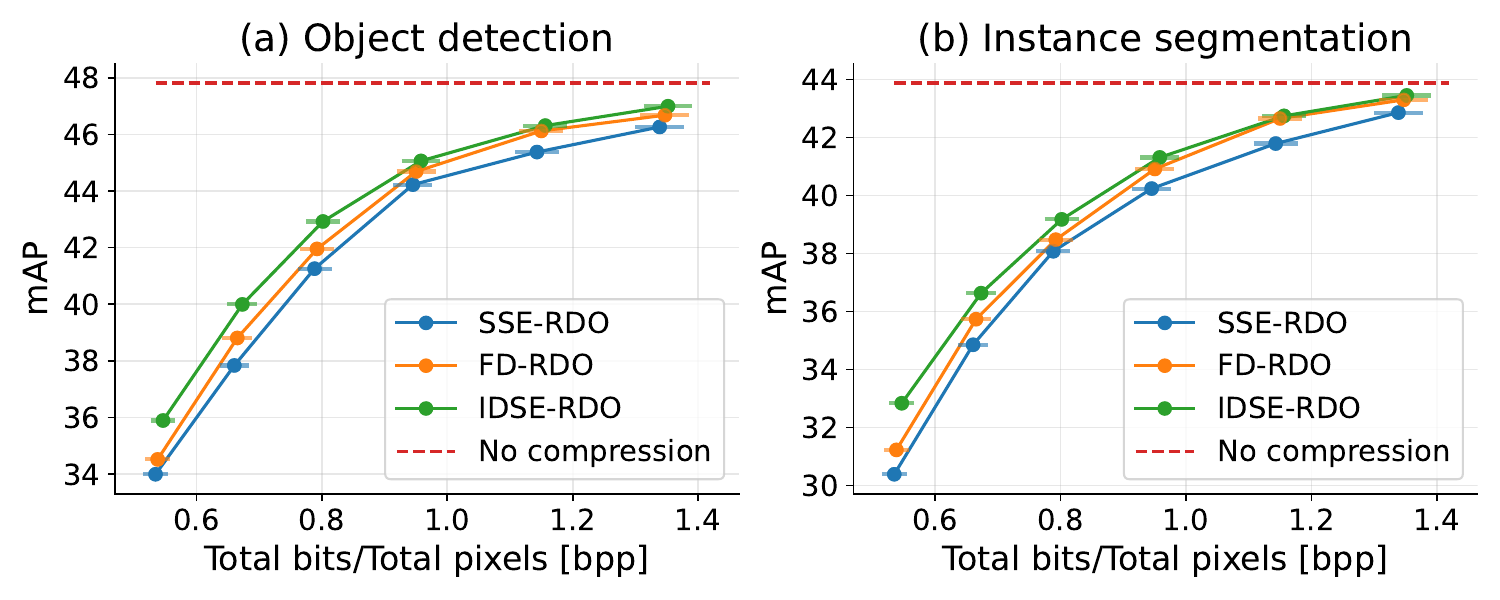}
    \includegraphics[width=0.49\linewidth]{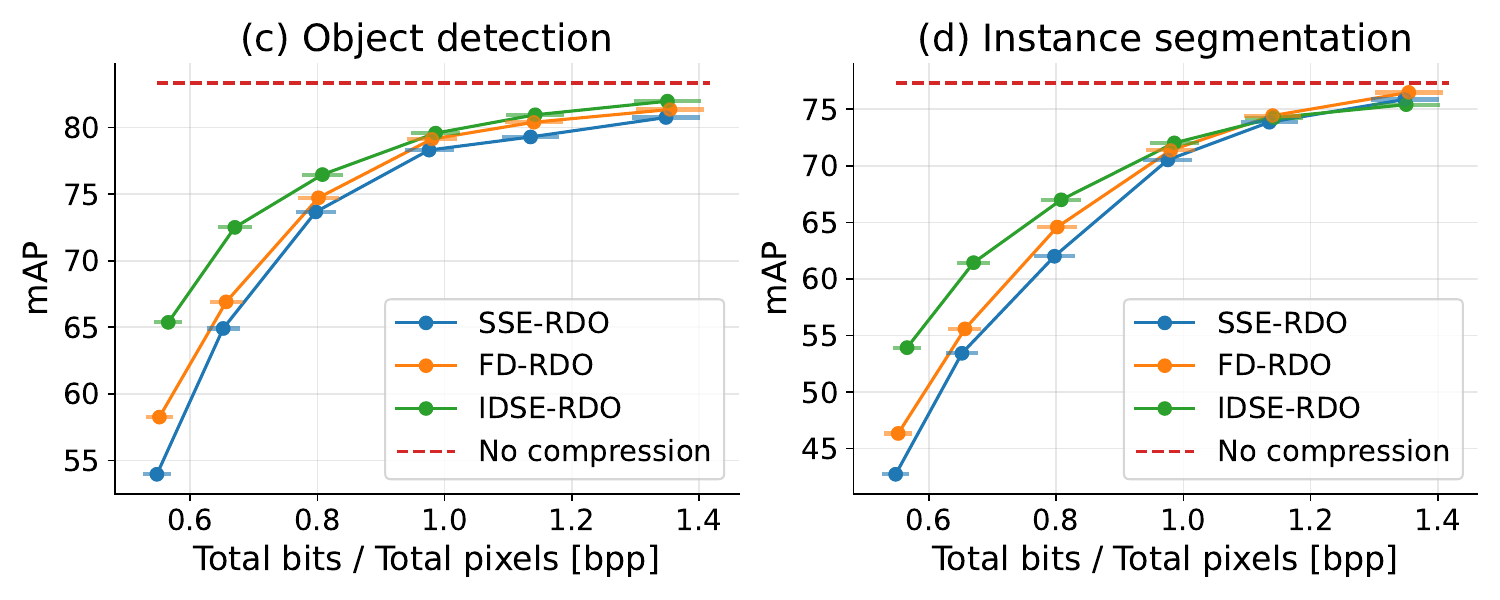}     
     \caption{RD curves for object detection and instance segmentation accuracy for AVC using SSE-RDO, our proposed IDSE-RDO with Rademacher sampling and $\ell = 8$ (IDSE-RDO), and RDO using the distance between features (FD-RDO) on $200$ images from the COCO dataset (a--b) and $50$ images from the PennFudan dataset (c--d). We also added the standard error on the estimation of the average bit-rate on top of each point, as a horizontal bar.}
     \label{fig:coco_res}
 \end{figure*}

We freeze the feature extractor and train the region proposal layers for $5$ epochs using a training set of $50$ images. We use the remaining $50$ images for testing. We provide the mAP-BD rate saving with respect to SSE-RDO AVC in Table~\ref{tab:metrics_results} (PF) and the RD curve for Rademacher ($\ell=8$) in Fig.~\ref{fig:coco_res}(c--d). IDSE-RDO, with the same feature extractor, also helps in this task. 

\begin{table}[t]
    \centering

    \begin{tabular}{lcccc}
        \toprule
         \textbf{Dimensions} & \textbf{Gaussian} [s] & \textbf{Rademacher} [s] & \textbf{DCT top16} [s] \\
        \midrule
            $\ell = 2$ & 0.079 & 0.067 & 0.072  \\	
            $\ell = 4$ & 0.120 & 0.112 & 0.123  \\	
            $\ell = 8$ & 0.241 & 0.212 & 0.231  \\	
        \bottomrule
    \end{tabular}    
    \caption{Average time to compute the Jacobian over $200$ images from the COCO 2017 validation set.}
    \label{tab:table_times}
\end{table}

\begingroup
	\begin{table}[t]
		\centering
		\setlength{\tabcolsep}{2.5pt} 
		\begin{tabular}{llccccc}
			\toprule & \textbf{Method}
			& \textbf{mAP seg.} [\%] & \textbf{mAP det.} [\%] & \textbf{PSNR} [\%] &
   \textbf{MS-SSIM} [\%] \\
			\midrule
   
            \multirow{12}{*}{\rotatebox{90}{\textbf{COCO}}} &
			$\ell = 2$ R. & $-6.01$ & $-6.31$ & $1.12$ & $-2.78$ \\	
			& $\ell = 2$ G. & $-6.19$ & $-6.11$ & $0.89$ & $-2.74$ \\	
			&$\ell = 2$ DCT & $-5.18$ & $-7.18$ & $1.84$ & $-1.82$ \\		
			\cmidrule{2-6}
			& $\ell = 4$ R. & $-7.06$ & $-7.18$ & $0.82$ & $-3.15$ \\	
			& $\ell = 4$ G. & $-6.81$ & $-7.24$ & $0.74$ & $-3.36$ \\	
			& $\ell = 4$ DCT & $-5.45$ & $-7.22$ & $1.82$ & $-2.19$ \\
			\cmidrule{2-6}
			& $\ell = 8$ R. & $\mathbf{-7.77} $ & $-8.28$ & $0.79$ & $-3.41$ \\	
			 & $\ell = 8$ G. & $-7.31$ & $\mathbf{-8.34}$ & $0.71$ & $\mathbf{-3.48}$ \\	
			&$\ell = 8$ DCT & $-5.47$ & $-8.21$ & $1.62$ & $-2.13$ \\
			\cmidrule{2-6}
            &FD & $-5.62$ & $-5.81$ & $\mathbf{0.66}$ & $\ \ 0.29$ \\	
			\midrule
            \midrule
            \multirow{4}{*}{\rotatebox{90}{\textbf{PF}}}
    		& $\ell = 8$ R. & $\mathbf{-9.09}$ & $\mathbf{-10.01}$ & $0.33$ & $\mathbf{-2.92}$ \\	
    		& $\ell = 8$ G. & $-8.85$ & $-9.26$ & $\mathbf{0.31}$ & $-2.87$ \\	
    		& $\ell = 8$ DCT & $-9.02$ & $-8.82$ & $0.39$ & $-2.71$ \\
			\cmidrule{2-6}
    		  & FD & $-4.34$ & $-4.96$ & $0.53$ & $0.52$ \\
    		\bottomrule
    		\end{tabular}		
		\caption{BD-rate saving with respect to SSE-RDO AVC, for $200$ images from the COCO 2017 validation set and $50$ images from the PennFudan (PF) dataset. R. stands for Rademacher, G. for Gaussian, FD for feature distance, seg.~for instance segmentation, and det.~for object detection. We keep $\ell$ features after dimensionality reduction. More negative is better. The best method for each metric is shown in boldface.}
		\label{tab:metrics_results}
    \vspace{-1em}
	\end{table}	 
 \endgroup

\section{Conclusion}
In this paper, we proposed a compression method that preserves the distance between the outputs of a feature extractor via RDO. Using linearization arguments and randomized dimensionality reduction, we simplified the distance between features to an input-dependent squared error loss involving the Jacobian of the feature extractor. This loss can be computed block-wise and in the transform domain. The Jacobian can be obtained before compressing the image, which provides computational advantages. We validated our method using AVC, which performs RDO to select between $4\times 4$ and $16\times 16$ prediction modes. Results show coding gains for computer vision tasks without significantly increasing the computing time. Future research will include extensions to account for saturation effects \cite{xiong_rate_2023} and more complex codecs.

\bibliographystyle{IEEEbib}
\bibliography{IEEEabrv,conference_101719}

\end{document}